# Non Destructive Determination Of Elastic Moduli By Two Dimensional Fourier Transformation And Laser Ultrasonic Technique

Xinya Zhang, Emmanuel Lafond and Ted Jackson
*Institute of Paper Science and Technology, at the Georgia Institute of Technology, 500 10$^{th}$ St., N.W., Atlanta, GA 30332*

Broadband laser ultrasonics and two dimensional Fourier transformation are used to characterize the properties of varieties of foils and plates. Laser ultrasonics generation is achieved by use of a pulsed laser which deposits pulsed laser energy on the surface of the specimen. The displacement amplitude of the resulting broadband ultrasonic modes are monitored using a two wave mixing photo-refractive interferometer. By applying a two dimensional Fourier transformation to the detected spatial and temporal displacement waveforms, the images of density of state (DOS) for the excited ultrasounds are obtained. Results are presented for a 150 μm thick paper sample, a 52.8 μm stainless steel foil and a 1.27 mm thick aluminum plate. The DOS image demonstrates the ability to measure the properties of each generated ultrasonic modes and provides a direct, non destructive, measure of elastic moduli of the tested specimens.

Techniques that have been used successfully for non-destructive elastic properties evaluation in films fall into two basic categories [1]: laser based ultrasonic methods [2-4] and techniques utilizing Brillouin scattering [5-7]. In laser ultrasonics (LU) experiments, ultrasonic waves are externally generated, and either the dispersion of surface propagation modes is measured or the acoustic pulses propagating out of the plane of the specimens are monitored. Brillouin scattering probes the thermally induced vibration modes and measures their dispersions.

The LU metrology uses pulsed laser irradiation to induce ultrasounds in a test object via three mechanisms: thermal-elastic, ablation and plasma. The resulting ultrasonic wave packets are often measured in a non-contact, nondestructive manner using an optical interferometer. LU techniques can provide information about the mechanical, thermal, and electronic properties of materials, and can also be used to detect surface and subsurface flaws in structures [2].

LU often monitors an ultrasonic waveform in the time domain. From the obtained time-domain waveform, a technique, like phase velocity calculation, can be used to obtain indirectly the phase velocity dispersion curve as a function of frequency[8-10]. Using a fit to this curve one can extract the physical properties of the sample. The conventional technique to obtain phase velocity dispersion requires skill, knowledge and know-how in data processing. In addition, previous knowledge of material properties is also necessary.

In this letter, we describe a two dimensional fast Fourier transformation (2D FFT) technique to avoid most problems confronted by the conventional method. It is implemented by taking incremental measurements in separation distance between ultrasound emitter and receiver to obtain a second spatial dimension, correspond to wavevector ( $k$ ) domain in Fourier transformation. The method automatically produces density of state ( DOS ) image as function of frequency ( $f$ ) and wavevector for the generated ultrasounds with minimum human involvement in the process. The DOS image not only beautifully displays all ultrasonic modes excited, but also separates them and provides a direct way to measure the properties for each mode such as dispersion, frequency range, modes displacement magnitude distribution, etc. By tracing the maximum of the DOS image, the phase velocity dispersions of individual modes can be directly obtained without previous knowledge of the sample properties.

The optical layout of our LU system is described in references [8-10]. Briefly, broadband ultrasonic waves are generated by a pulsed, point focused laser beam that is obtained from a 1064 or 532 nm Nd:YAG laser. The resulting ultrasonic waves propagating through the sample are detected by an interferometer using the beam of a CW coherent laser. The ultrasonic packets induce displacements at the specimen surface which modulate the phase of the reflected CW laser beam. The unprepared sample surfaces are often rough, and subsequently produce multiple speckles, which can make the laser detection of ultrasounds challenging. A photorefractive interferometer [8] is used for this purpose. Photorefractive interferometers are multiple speckle interferometers that collect light from many reflected speckles by adapting the wavefront of a reference beam to the irregular wavefront of the signal beam coming from the sample. They provide better sensitivity to ultrasonic displacements than a homodyne Mach-Zehnder or Michelson interferometer, when used on light scattering samples [8].

The samples tested are commercially available copy papers, stainless steel (type 304) foils, and aluminum plates. No special preparation was needed for the sample surface. The pulsed laser energy is about 3 mJ/pulse on the copy paper samples and 6 mJ/pulse on metal samples. The generated waveforms are collected and digitized by a Gage computer acquisition board under LabVIEW environment, the 2D FFT are executed through LabVIEW built-in functions.

In this letter, results of only three samples, i.e., ~ 150 μm thick copy paper specimen, 52.8 μm thick stainless steel foil and 1.27 mm aluminum plate are presented.

Fig 1 illustrates the 2D FFT image for a copy paper sample along the machine direction (MD), the direction along which the wood fibers are aligned. It is evident that the image not only beautifully displays different excited ultrasonic modes, but also provides information for each mode such as dispersion, frequency range, modes displacement magnitude distribution.

There are two dominant features in the image. The mode with dispersion following a straight line is a sound wave propagating through air (shock waves). This wave is generated during the plasma regime of ultrasound generation, and detected by the photo-refractive interferometer through the index modulation of the air between the detector and the sample as the shock wave travels through. This is an unwanted feature in laser ultrasonic generation, but somehow we can not avoid its existence.

The second feature is A0 wave (the lowest order asymmetric Lamb wave ). It is very dispersive and has higher displacement magnitude at lower frequency. As the frequency increases, its dispersion behaves more and more like a straight line and its magnitude decreases rapidly. But one can see clearly that up to 900 KHz, the A0 wave is still detectable. As for the low frequency limit, which is subject to overlap with shock sound waves and environmental vibration noise, it is easy to distinguish the A0 wave down to 20 KHz.

In LU, the dispersion curves of $A_0$ waves is an important channel to extract material properties. For the copy paper specimen, two important parameters for industrial processing: flexural rigidity (FR) and shear rigidity (SR), can be obtained from the A0 dispersion curve. The extracted FR ( 1.8E-3 N·m ) and SR ( 3.6E4 N/m ) for this paper sample along the MD and measurements on the other paper specimens are consistent with the contact ultrasonics measurements [11] and previous laser ultrasonics measurements [10].

The 2D FFT image not only demonstrates reliably the A0 dispersion curves, but separates the shock sound wave traveling through air from A0 excitation as well. It is evident in Fig. 1 that A0 and shock sound waves are present, and are well separated in the 2D FFT image. The presence of shock waves has no side effect on the behavior of Ao wave. This is an another noteworthy advantage of the method, because phase velocities of both waves are relatively similar in copy paper specimens at low frequencies. Also since both modes have a relatively high power density over the frequency range of interest, it is not an easy task to separate the two waves by the conventional phase calculation methods or other technique.

Fig 2 represents a 2D FFT image for a 52.8 μm thick stainless steel foil. The metal is a better media for high frequency ultrasounds as its very small grain size does not diffract ultrasonic waves as much as wood fibers. The A0 wave has a considerable DOS magnitude up to 1.7 MHz. The shock wave traveling through air is also present in the image, from DC up to 1.2 MHz. There is a third curve in the frequency range of 1.2MHz to 1.7 MHz, which represents the contribution from the reflected A0 wave from one edge of the sample.

The third curve in Fig. 2 exhibits another advantage of this technique, i.e., to separate the reflected $A_0$ waves by the sample edges or sample holder. Reflection of a Lamb mode from an edge is quite a complicated problem[12-13]. In time domain waveforms, their contributions are displayed and overlap with $A_0$ displacement curves. To simplify the data processing, a smoothing procedure is often utilized to reduce their influences. Nevertheless they result in anomalies and disturbances in FT magnitude and phase diagrams in some circumstances, and cause inaccuracies and misleads in the conventional phase velocity calculations. In the 2D FFT images, the reflected waves may result in extra lines or curves, however they have no side effect on the dispersion of the direct excited modes. This had been further proven experimentally by comparing images taken at the center and close to one edge of the sample (images not shown in this letter), and the obtained dispersion curves had a negligible difference.

The 2D FFT images that provide a reliable A0 dispersion curve therefore enable to determine Young's modulus and shear modulus for those isotropic foils. The obtained data are in a good agreement with the products specifications of the foil provider. For elastically anisotropic specimens like papers, by rotating samples around the normal to the surface, the images can provides the information necessary to extract the related direction oriented properties. This makes this technique a powerful and effective tool.

Figure 3 (a) shows one 2D FFT image for a 1.27 mm thick aluminum plate. The image not only shows the dispersion curve for A0 modes, but S0 (the lowest order symmetric Lamb mode) and higher order Lamb modes [14-15] as well. As expected, for low frequencies, both A0 and S0 are well separated from each other. As the frequency increases, A0 and S0 merge and essentially become a non-dispersive Rayleigh surface mode localized at the sample surface. Higher order lamb modes are standing modes that resulted from the interference of waves reflected at two free boundaries of the plate. They are classified as longitudinal ( L ) or transverse (T) characterized standing modes [15]. For this sample, L1, L2, T1, T2 are successfully observed .

Figure 3(b) illustrates the calculated two dimensional image of density of state (DOS) for the excited modes through a Green's function formalism [16] as functions of wave vector and frequency. The brightness of the image varies directly as the ultrasounds DOS associated with the $G_{33}$ component. It is this component, perpendicular to plate surface, that evaluates the magnitude of 2D FFT by LUS. The best fits (Fig. 3(b)) were obtained using a set of elastic constants for the Aluminum plate ($C_{11}$= 117.5 GPa and $C_{44}$ = 25.1 GPa ) and the known density (2.69 g/cm$^3$)[17]. It is evident from Fig. 3 that the DOS image not only predicts the dispersions of all observed modes, but reveals a good agreement of the magnitude of the generated ultrasounds as well.

One can notice that the magnitudes of higher order Lamb modes are weak and modes like T3, T4. L3 etc, are absent in the experimental image. That is partially accounted for by the pulsed laser generation method, for which the magnitude of the generated ultrasound is exponentially decreasing with frequency. This is also partially accounted for by the small amplitude or complete lack of an out of plane displacement component, which is the component the interferometer is sensitive to. The later is also demonstrated in DOS image in Fig 3(b), where only the projected DOS is illustrated.

The L, T excitations are respectively most sensitive to the stiff elastic constant $C_{11}$ and shear elastic constant $C_{44}$ of the Aluminum plate. Hence those two principal elastic constants are extracted from L and T ultrasounds dispersion curves. The $C_{ij}$ elastic parameters ($C_{11}$= 117.5 GPa and $C_{44}$ = 25.1 GPa) deduced from the image illustrate the ability of this technique to evaluate the elastic parameters of the plate in a non-contact and non-destructive way. The elastic constants yield the Young's (E = 68.5 GPa) and shear (G = 25.1 GPa) moduli for the aluminum plate. The values for the moduli are in a good agreement with literature values ( E = 70.3 GPa and G = 26.1 GPa ) [17].

In conclusion, a 2D FFT image method is introduced for laser ultrasonics. The image provides a rich information including the dispersion of the generated modes, the modes density of state, and their frequency range. Copy paper samples, stainless steel foils and aluminum plates were tested. For the thin paper specimen and steel foil, the 2D FFT images for the $A_0$ waves are displayed. For the aluminum plate, different high order Lamb modes as well as $A_0$ and $S_0$ were presented in the image. The many advantages over the conventional method were detailed and demonstrate this technique is a powerful and effective tool for the non-contact, non-destructive testing of membranes, foils and plates ranging from tens of μm to several mm thick.

This work was supported by an AMRC grant (internal funding) from the Institute of Paper Science and Technology at the Georgia Institute of Technology. We are grateful of Prof A.G. Every for useful discussions.

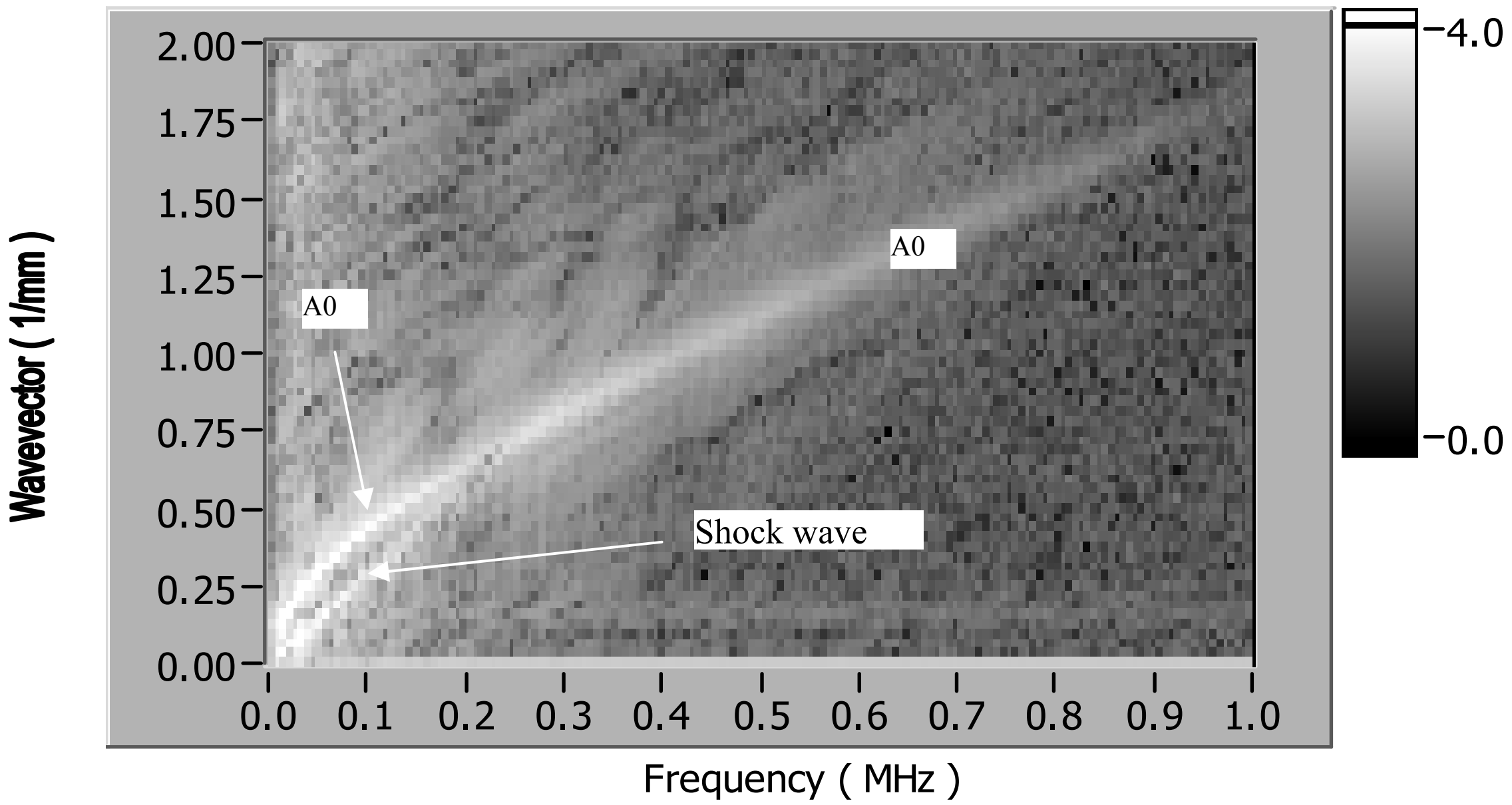

Fig 1 : 2D FFT image for a copy paper sample. The mode with a dispersion following a straight line is the shock sound wave that travels through air. The A0 (the lowest order asymmetric Lamb mode) is dominant in a frequency range of 20 KHz up to 0.9 MHz.

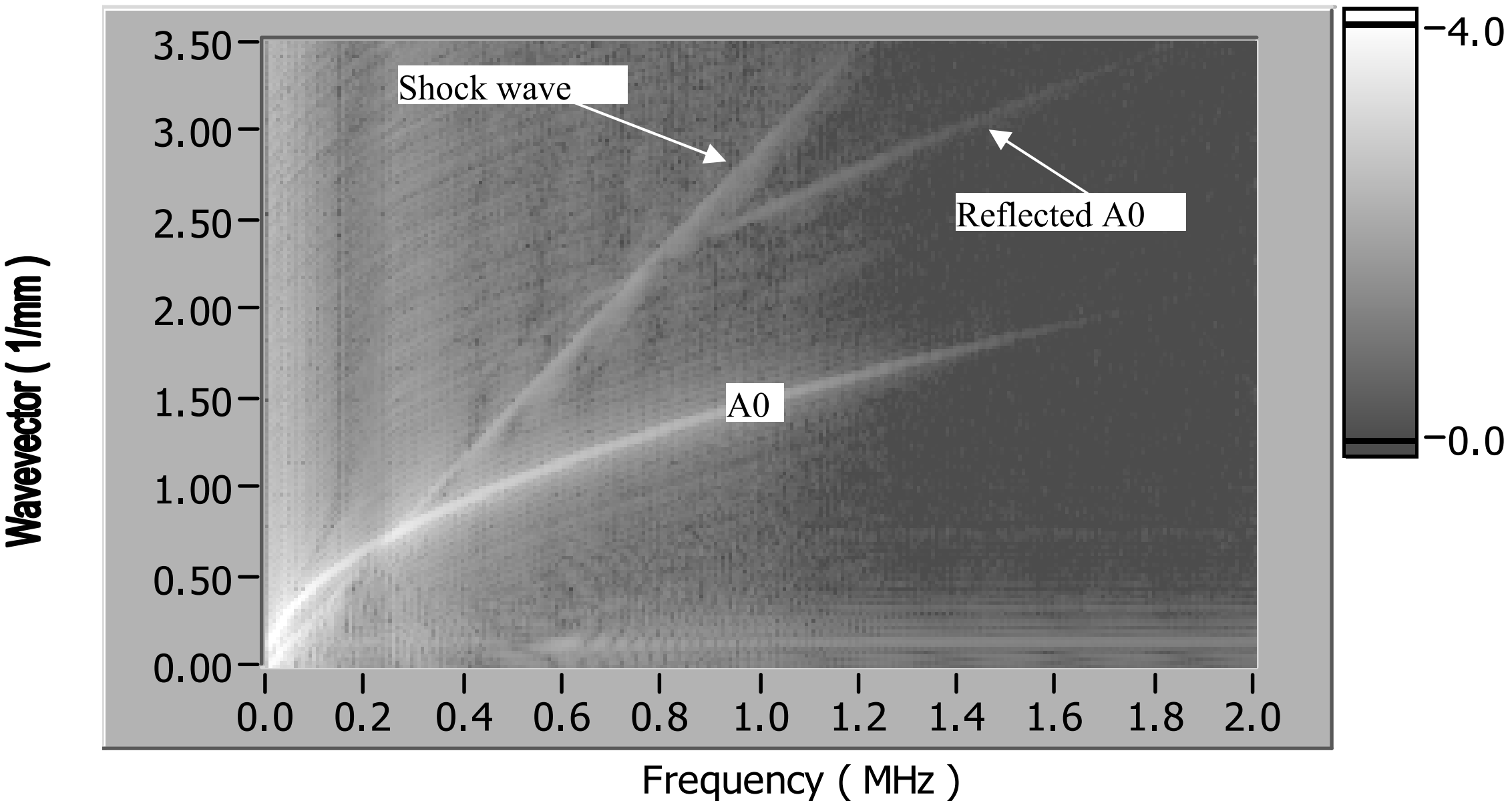

Fig. 2, 2D FFT image for a 52.8 µm thick stainless steel foil. A0 is the lowest order asymmetric Lamb modes. Reflected A0 comes from the reflection of A0 from the sample holders or sample edges. The shock wave is a sound wave traveling through air.

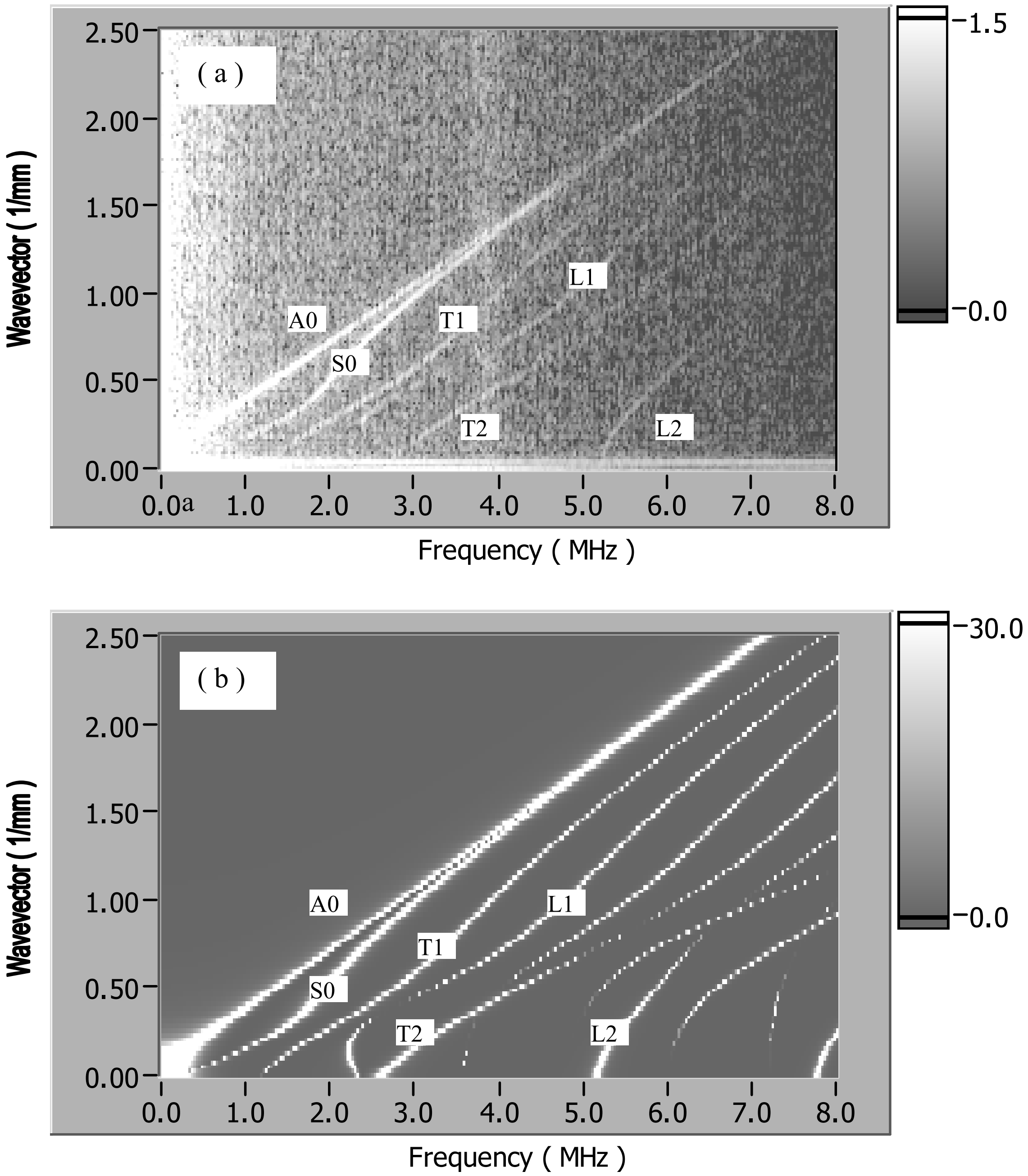

Fig 3 2D FFT image (a) and calculated DOS image (b) for a 1.27 mm thick aluminum plate. A0, S0 are respectively the lowest order asymmetric and symmetric Lamb modes. L, T are higher order longitudinal, transverse characterized Lamb modes.